\documentclass[reprint, superscriptaddress, amsmath, amssymb, aps, physrev]{revtex4-2}

\usepackage{graphicx}

\usepackage[dvipsnames,table]{xcolor}
\usepackage[colorlinks=true,linkcolor=RoyalBlue,citecolor=Brown,urlcolor=Brown,filecolor=Brown]{hyperref}

\usepackage{booktabs,todonotes,upgreek}

\makeatletter
\def\CT@@do@color{%
  \global\let\CT@do@color\relax
        \@tempdima\wd\z@
        \advance\@tempdima\@tempdimb
        \advance\@tempdima\@tempdimc
\advance\@tempdimb\tabcolsep
\advance\@tempdimc\tabcolsep
\advance\@tempdima2\tabcolsep
        \kern-\@tempdimb
        \leaders\vrule
                \hskip\@tempdima\@plus  1fill
        \kern-\@tempdimc
        \hskip-\wd\z@ \@plus -1fill }
\makeatother

\begin{document}

\title{\textbf{High-Efficiency Quantum-State Detection of ThF$^+$ with Resonance-Enhanced Multiphoton Asymmetric Dissociation} 
}

\author{Kia Boon Ng}
\email{kbng@triumf.ca}
\affiliation{TRIUMF, Vancouver, British Columbia, V6T 2A3, Canada}
 
\author{Sun Yool Park}
\affiliation{JILA, NIST and University of Colorado, and Department of Physics, University of Colorado, Boulder CO 80309, USA.}

\author{Anzhou Wang}
\affiliation{JILA, NIST and University of Colorado, and Department of Physics, University of Colorado, Boulder CO 80309, USA.}

\author{Addison Hartman}
\affiliation{JILA, NIST and University of Colorado, and Department of Physics, University of Colorado, Boulder CO 80309, USA.}

\author{Patricia Hector Hernandez}
\affiliation{JILA, NIST and University of Colorado, and Department of Physics, University of Colorado, Boulder CO 80309, USA.}

\author{Rohan Kompella}
\affiliation{JILA, NIST and University of Colorado, and Department of Physics, University of Colorado, Boulder CO 80309, USA.}

\author{Lan Cheng}
\affiliation{Department of Chemistry, The Johns Hopkins University, Baltimore, MD 21218, USA}

\author{Stephan Malbrunot-Ettenauer}
\affiliation{TRIUMF, Vancouver, British Columbia, V6T 2A3, Canada}
\affiliation{Department of Physics, University of Toronto, Toronto, Ontario, M5S 1A7, Canada}

\author{Jun Ye}
\affiliation{JILA, NIST and University of Colorado, and Department of Physics, University of Colorado, Boulder CO 80309, USA.}

\author{Eric A. Cornell}
\affiliation{JILA, NIST and University of Colorado, and Department of Physics, University of Colorado, Boulder CO 80309, USA.}

\date{\today}

\begin{abstract}
Efficient quantum-state detection is crucial for many precision control experiments, such as the ongoing effort to probe the electron's electric dipole moment using trapped molecular $^{232}\mathrm{ThF}^+$ ions at JILA. While quantum state detection through state-selective photodissociation has been successfully implemented on this molecule, progress has been hindered by low dissociation efficiency. In this work, we perform spectroscopy on the molecule to identify excited states that facilitate more efficient photodissociation. For the most favorable transition, we achieve a dissociation efficiency of 57(14)\% with quantum state selectivity. Additionally, we discuss several state detection protocols that leverage favorable excited states that will facilitate simultaneous readout of all EDM relevant states, allowing further improvement of overall statistics.
\end{abstract}

\maketitle

\section{Introduction}

    The field of quantum state control of molecules has seen remarkable growth over the past decade, leading to groundbreaking advancements in quantum information, quantum simulation, the probing of chemical reactions at the quantum level, and precision measurements of new physics \cite{langen2024quantum, demille2024quantum, ye2024essay}. Our current work builds on an effort to refine state-selective readout in molecules, paving the way for precision measurements that explore physics beyond the Standard Model (BSM) of particle physics. One manifestation of BSM physics is the presence of electric dipole moments (EDMs) in both elementary and composite particles \cite{POSPELOV2005119, ENGEL201321, Chupp2019electric, cesarotti2019interpreting}. Measurements of EDMs, e.g., Refs.\ \cite{Roussy2023an, acme2018improved, hudson2011improved, eckel2013search, zheng2022measurement, Sachdeva2019new, Graner2016reduced, bishof2016improved, regan2002new, murthy1989new, Cho1991search, nedm2020, Ramsey1950on}, serve to constrain the nature of new physics and guide us toward a more comprehensive understanding of the universe.
    
    Recent advancements in the quantum state control of trapped HfF$^+$ at JILA have resulted in the current best limit on the electron's EDM \cite{Roussy2023an, caldwell2023systematic}. Building on these efforts, the JILA group is now transitioning to $^{232}$ThF$^+$, which offers greater sensitivity to the electron's EDM than HfF$^+$ \cite{ng2022spectroscopy, gresh2016broadband, meyer2008prospects, Denis2015, skripnikov2015theoretical}. 
    Concurrently, TRIUMF has identified the radioactive isotopologue $^{227}$ThF$^+$ as a promising candidate for measuring the nuclear Schiff moment \cite{chen2024relativistic,flambaum2019enhanced,Liang1995level,hammond2002observation,minkov2024skyrme}. The nuclear Schiff moment manifests as a molecular EDM, which is sensitive to new physics within the atomic nucleus, further advancing the growing field of research on radioactive molecules \cite{Arrowsmith-Kron2024opportunities, garcia2020spectroscopy, udrescu2024precision, isaev2010laser, klos2022prospects, marc2023candidate, athanasakis2021radioactive}. Efficient and effective quantum state control of ThF$^+$, a less-studied molecule, is essential for the successful execution of these experiments.

    While spectroscopy had been conducted on ThF$^+$ \cite{barker2012spectroscopic, gresh2016broadband, zhou2019visible, zhou2020second, ng2022spectroscopy}, the overall performance of quantum state control for this molecule left much to be desired. Progress in the JILA ThF$^+$ experiment had been hindered by low efficiency in state readout techniques; in particular, resonance-enhanced multiphoton dissociation (REMPD) \cite{zhou2020second} achieved an estimated dissociation efficiency of only 4\%. In contrast, the dissociation efficiency in the JILA HfF$^+$ experiment was estimated to be about 20\%. Both the JILA and TRIUMF experiments stand to gain significantly from a state detection protocol that offers improved efficiency. This work will focus on identifying transitions within the molecule to enhance dissociation efficiency.

    This paper proceeds as follows. We present an overview of the energy landscape of ThF$^+$ in Section \ref{sec:energylandscape}. We give a summary of a protocol that enables two-state detection in Section \ref{sec:REMPAD}, with which we derive general guiding principles to search for intermediate states suitable for this purpose. We present our setup for the survey spectroscopy in Section \ref{sec:SS_setup}, followed by our methodology for extracting molecular parameters of rotational bands found in Section \ref{sec:SS_fitting}. We briefly discuss the complexity of the energy landscape above molecular dissociation threshold in Section \ref{sec:SS_landscape}. We show examples of asymmetric photofragments and discuss state selectivity and dissociation efficiency for select bands in Section \ref{sec:O2}. In Section \ref{sec:O0}, we explore the prospects of using $\Omega=0^\pm$ intermediate states for multi-state detection. Finally, we conclude with Section \ref{sec:conclusion}.

\section[Energy Landscape of ThF+]{Energy Landscape of ThF$^+$}\label{sec:energylandscape}

\begin{figure}[h!]
        \centering
        \includegraphics[width=\columnwidth]{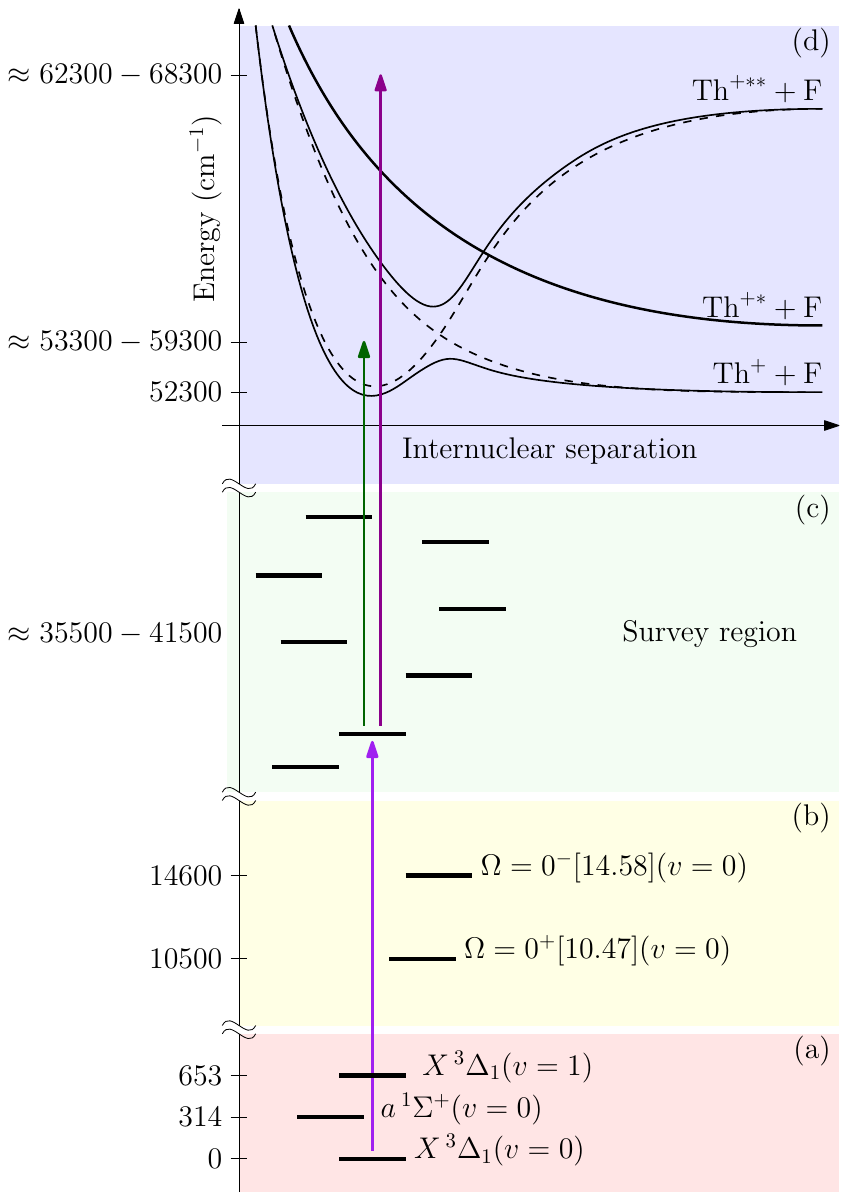}
        \caption{\textbf{Schematic overview of the energy landscape of ThF$^+$.} Not drawn to scale. The numbers along the vertical axis indicate energy above the ground state. The horizontal axis in panel (d) corresponds to the internuclear separation of the atoms within the molecule. The horizontal axis does not convey meaning in other panels. Horizontal lines in panels (a-c) indicate energy levels. (a) Lowest-lying states. EDM measurements are performed in the $X\,^3\Delta_1(v=0)$ vibronic manifold. (b) Low-lying electronic states used for optical pumping. Numbers in square brackets correspond to the energy of the states in units of 1000~cm$^{-1}$. (c) High-lying electronic states used for REMPD. (d) Conceptual simplified representation of the pre-dissociating energy landscape. The dissociation energy is 52300~cm$^{-1}$. Dashed lines represent diabatic curves connecting to different asymptotic atomic states. Th$^{+*}$ and Th$^{+**}$ represent two (of many) excited states of the thorium ion. In this example sketch, non-adiabatic couplings can result in avoided crossings, resulting in adiabatic curves illustrated by two of the black solid lines. In this work, we perform spectroscopy to search for states in panel (c). The arrow connecting states in panels (a) and (c) indicate a tunable first REMPD photon. The arrows connecting panels (c) and (d) indicate the second REMPD photons. We repeat the spectroscopy with two different energies for the second REMPD photon to access different energy regions in panel (d) (see Section \ref{sec:SS_setup} in the main text).}
        \label{fig:energylandscape}
    \end{figure}

    The energy landscape can be largely split into four regions, as indicated by the four colored panels labeled (a) to (d) in Figure \ref{fig:energylandscape}. 
    
\begin{figure}[htb]
        \centering
        \includegraphics[width=0.72\columnwidth]{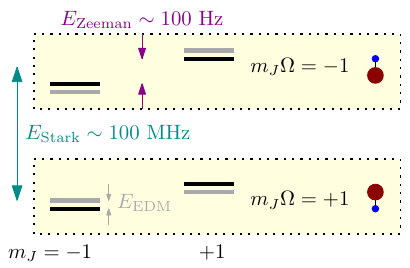}
        \caption{\textbf{Schematic diagram of the EDM-sensitive states within the $X\,^3\Delta_1(v=0,J=1)$ manifold.} Not drawn to scale. Hyperfine structure is omitted and only the EDM-sensitive states are shown for clarity. Within this manifold, there are two Stark doublets, each described by quantum numbers $m_J\Omega = +1$ and $-1$, corresponding to the molecule aligned parallel and anti-parallel to the polarizing electric field, respectively; the electric and magnetic fields point upward in this figure. The doublets are split by the Stark shift, and the energy levels within each Stark doublet are split by the Zeeman shift. EDMs introduce additional shifts in the energy levels, resulting in energy levels indicated by the gray lines.}
        \label{fig:sciencestates}
    \end{figure}

    We use the ground state in ThF$^+$, $X\,^3\Delta_1(v=0,J=1)$ to measure the EDM, where $v$ and $J$ correspond to the vibrational and rotational quantum numbers, respectively. In the presence of a bias electric field with a magnitude of $\gtrsim 5~\mathrm{V/cm}$, within the $X\,^3\Delta_1(v=0,J=1)$ ground rovibronic manifold, there are two Stark doublets, described by quantum numbers $m_J\Omega = +1$ and $m_J\Omega=-1$, corresponding to the molecule aligned parallel and anti-parallel to the external electric field, respectively.
    Here, $\Omega$ and $m_J$ are the projections of $J$ onto the internuclear axis and lab frame, respectively. These levels are illustrated in Figure \ref{fig:sciencestates}.
    
    These Stark doublets have very similar magnetic field susceptibilities \cite{ng2022spectroscopy} and have sensitivities to the EDM with opposite signs. An EDM measurement is, in its essence, a measurement of the difference of the energy splittings of each Stark doublet \cite{caldwell2023systematic}.
    
    We can move the population of our molecules between states in panel (a) of Figure \ref{fig:energylandscape} using optical pumping through intermediate states in panel (b). In this work, we survey many potential intermediate states, as in panel (c) of Figure \ref{fig:energylandscape}, with the goal of improving REMPD-based readout of the population in the various states in panel (a).
    
\section{Resonance-Enhanced Multiphoton Asymmetric Dissociation}\label{sec:REMPAD}

    The use of resonance-enhanced multiphoton asymmetric dissociation (REMPAD) for two-state readout has seen great success in HfF$^+$ \cite{Roussy2023an, caldwell2023systematic}. Our two-state readout protocol allows us to simultaneously read out the population in states of a selected $m_J$ in both Stark doublets. This enables us to perform EDM measurements on both Stark doublets simultaneously, bringing two benefits: (i) enabling faster collection of ion numbers for better statistics, and (ii) suppressing common-mode noise from ion production and magnetic field fluctuations. For more information, please refer to Ref.\ \cite{caldwell2023systematic} and references therein.
    
    The same technique has also been demonstrated in ThF$^+$ \cite{zhou2020second} but with lower efficiency \cite{ng2023thesis}. We set out to search for new intermediate states with suitable quantum numbers to boost the efficiency of the REMPAD protocol. 
    
    The physics behind maximizing the efficacy of REMPAD is rather involved (see, e.g., Refs. \cite{beswick2008quantum,mcdonald2016photodissociation,majewska2018experimental,ng2023thesis}). Here, a brief summary follows. 
    
\begin{figure}[htb]
        \centering
        \includegraphics[width=\columnwidth]{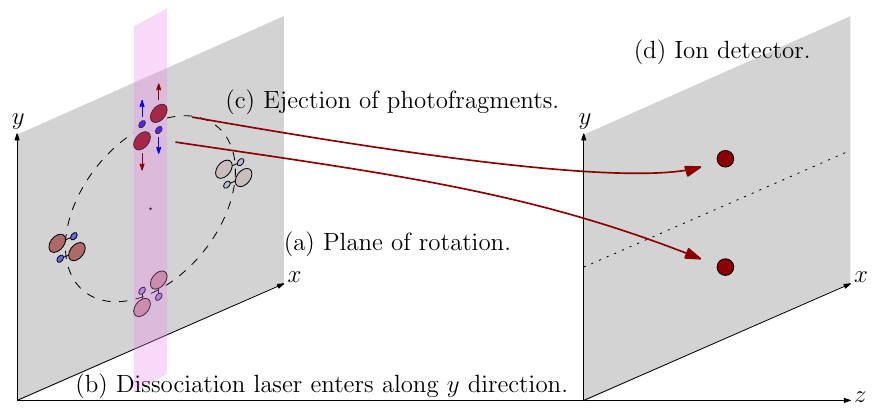}
        \caption{\textbf{Schematic diagram of relative directions of dissociation laser beam, rotating electric field, and ion detector.}         
        (a) A rotating electric field in the $x$--$y$ plane defines the direction of molecular orientation. ThF$^+$ molecules are prepared in a statistical mixture of states aligned either parallel or anti-parallel to the field. The molecular orientations are depicted with varying transparency to represent snapshots in time, as they adiabatically follow the instantaneous direction of the rotating field. At each snapshot, the rotating electric field points towards the dot at the center of the rotating micromotion.
        (b) Laser beams (shown in pink) are introduced along the $y$ axis at the moment when the molecular ensemble is instantaneously aligned with this direction. At this instance, the rotating electric field is pointing in the $-\hat{y}$ direction. Photofragments are emitted along the $\pm y$ direction, depending on the orientation of each molecule at the time of dissociation.
        (c) After dissociation, the rotating field is switched off and an electric field along the $z$ axis ejects the Th$^+$ ions toward an ion detector positioned along the same direction.
        (d) The $(x, y)$ impact positions of the Th$^+$ ions on the detector encode their velocities in the $x$--$y$ plane. Ions ejected in the $+y$ direction during dissociation land on the $+y$ side of the detector, and \textit{vice versa}.}
        \label{fig:PhotofragmentEjection}
    \end{figure}

    In the presence of an external electric field, heteronuclear molecules like ThF$^+$ can have well-defined orientations (i.e., parallel or anti-parallel) with respect to the field, as shown in Figure \ref{fig:sciencestates}. In the context of an EDM measurement with ThF$^+$, the quantum states of interest exhibit molecular orientations that are maximally aligned and anti-aligned with the field. Semiclassically, the photofragments (i.e., Th$^+$ and F) are ejected along their alignment direction upon dissociation. By directing the ejected ions towards an ion detector in a direction orthogonal to the direction of photofragment ejection as seen in Figure \ref{fig:PhotofragmentEjection}, we can identify ions landing on one half of the detector as belonging to one quantum state and those on the other half as belonging to another quantum state. 
    
    Quantum mechanics dictates that the angular distribution of ejected photofragments depends on both the quantum state of the molecule before dissociation and the photon used for dissociation. Angular distributions that are highly directional enhance the contrast in two-state detection using REMPAD. To achieve maximal directionality, it is generally advantageous to dissociate from a state where the quantum numbers satisfy $J = |\Omega| = |m_J|$, as this maximizes molecular alignment with the lab frame. For our purposes, it also helps to work with the largest accessible $J = |\Omega| = m_J$ state \cite{ng2023thesis}. 

    When working with a thermal cloud of ions, it is essential that the kinetic energies of the photofragments are predominantly due to the dissociation process itself, with smaller contributions from the thermal energies that the ions possess before dissociation. An appealing approach might be to excite the molecules to energies significantly above the dissociation threshold to increase the dissociation kinetic energy. However, because thorium is a heavy atom, there exists a complex landscape of dense molecular potential energy curves (hinted at panel (d) in Figure \ref{fig:energylandscape}) above the dissociation threshold. As a result, there is a risk of transitioning to a molecular potential curve that asymptotically connects with excited states of Th$^{+}$ and F. In such cases, excess energy above the dissociation threshold may go into exciting the atomic internal degrees of freedom of Th$^{+}$ and F, rather than being released as kinetic energy in the photofragments.

    The EDM-sensitive state in ThF$^+$ is the $X\,^3\Delta_1 (J=1, m_J=\pm 1)$ state, i.e., $J = |\Omega| = |m_J| = 1$. Hence, we wish to drive a transition with a circularly polarized photon (the first REMPD photon) on an R(1) transition to a $J = |\Omega| = |m_J| = 2$ intermediate state. The second REMPD photon, also circularly polarized, subsequently excites the molecule to an energy above the dissociation threshold. The general guiding principles for the survey spectroscopy are thus to focus on the search for $|\Omega|=2$ intermediate states with high dissociation efficiency and high photofragment kinetic energies. 

\section{Survey Spectroscopy}

    \subsection{Setup}\label{sec:SS_setup}
    
        The setup used for the REMPD spectroscopy is detailed in previous work, see Ref.\ \cite{ng2022spectroscopy} and references therein and Section 3.3 of Ref.\ \cite{ng2023thesis}. 
        Here is a brief summary. 
        
        After a molecular creation process including ablation, supersonic expansion, and state-selective photoionization \cite{zhou2019visible}, a thermal cloud of approximately 5000 $^{232}$ThF$^+$ molecular ions distributed across the $X\,^3\Delta_1(v=0,J=1-8)$ manifold is trapped in a 2D Paul trap with axial confinement (secular frequencies of about 2~kHz in all three directions). We can optionally populate the excited vibrational manifolds $X\,^3\Delta_1(v=1,J=1-2)$ or $a\,^1\Sigma^+(v=0)$ by applying optical pumping through states in panel (b) of Figure \ref{fig:energylandscape}. Two pulsed laser beams are introduced into the trap from the radial direction, with their $k$ vectors aligned in parallel, to dissociate the molecules. The pulsed lasers are represented by vertical arrows connecting panels (a) to (c) and (c) to (d) in Figure \ref{fig:energylandscape}. Although the transverse beam profiles of the pulsed lasers are highly multimodal, we make an effort to mode-match the beam envelopes with the $\sim 5~\mathrm{mm}$ rms size of the ion cloud.
        
        The first REMPD photon is generated by a dye laser with a 2400~l/mm grating, giving a linewidth of about 0.06~cm$^{-1}$ at 570~nm. The dye laser is pumped at 355~nm by a flashlamp-pumped Nd:YAG laser (10~ns pulse width, 10~Hz repetition rate) with a conversion efficiency of about 15\%. The output of the dye laser is frequency-doubled with a temperature-controlled BBO crystal. Dyes used in this work include Coumarin 153/540A and Coumarin 307/503. The wavelength of the frequency-doubled laser is swept across the entire region covered by the two dyes, from 35500~cm$^{-1}$ to 41500~cm$^{-1}$. 
        
        The second REMPD photon is generated by the selfsame pump laser that pumps the dye laser, using either the 532~nm or 355~nm harmonic. The second REMPD laser pulse, also 10~ns in duration, lags behind the first REMPD laser pulse by about 2~ns. Temporal delay between the two pulses can be adjusted with a delay line. 

        No polarizing electric field is used for the initial survey spectroscopy. For the follow-up studies characterizing angular distribution of photofragments through REMPAD, we apply a rotating electric field to polarize the molecules. For REMPAD studies, we use optical pumping to prepare our molecules in the $X\,^3\Delta_1(v=0,J=1,m_J=+1)$ state at an electric field strength of 60~V/cm, and reduce it to 20~V/cm for dissociation and efficient ion kickout.
        
        Immediately after the REMPD lasers hit the ion cloud, all the ions are ejected towards an ion detection setup formed by a microchannel plate assembly with phosphor screen (P43). The ejection direction is orthogonal to the plane of the rotating electric field and the $k$ vectors of the REMPD photons. Th$^+$ generated from the dissociation and the remaining ThF$^+$ are separated during the time of flight (about 180~$\upmu$s) and arrive at the ion detection setup separated by 12~$\upmu$s. The phosphor screen is time-gated to image the transverse positions of only ions arriving within the time window corresponding to that of Th$^+$. The slower moving ThF$^+$ ion bunch is not imaged, but the total charge collected during the ThF$^+$ window provides a useful shot-to-shot normalization \cite{shagam2020continuous}.

        For survey spectroscopy, the frequency of the first REMPD photon is scanned in steps of 0.5~cm$^{-1}$. The pulse energies used are 70~$\upmu$J to 400~$\upmu$J for the first REMPD laser pulse (lower energy towards the edge of the effective ranges of the dyes) and 25~mJ for the second REMPD laser pulse. Both REMPD photons are circularly polarized with the same handedness, unless otherwise stated. The entire scan for a combination of one dye and one choice of Nd:YAG harmonic for the second photon takes about 50~hours.

        We identify several strong bands from the initial survey spectroscopy and perform scans with finer steps (0.05~cm$^{-1}$ or 0.1~cm$^{-1}$) with reduced pulse energies to resolve the lines within each band for fitting purposes. Wavelengths of the first REMPD photon are measured with a wavemeter calibrated to the rubidium D2 line during the scan. 

    \subsection{Fitting to Rotational Bands}\label{sec:SS_fitting}

        Our survey of 6000~cm$^{-1}$ bandwidth reveals hundreds of rotational bands. Data is available in Ref.\ \cite{data}. We have the signal-to-noise and patience to perform finer scans for 57 of them. Molecular parameters for the bands are determined by fitting with the following procedure:
        \begin{enumerate}
            \item All observed bands conform to Hund's case (c). The energy of each observed line is fit to the following equation:
                \begin{equation}
                    E = T_0 + B' J' (J'+1) - B'' J'' (J''+1),
                \end{equation}
                where $T_0$ is the band origin of the excited state, $B$ is the rotational constant, $J$, once again, is the quantum number for the total angular momentum of the molecule, and singly (doubly) primed variables correspond to those of the excited (ground) electronic state.
            \item For an unsaturated transition, the intensity of each line is proportional to the initial population of each rotational state and by the H\"onl-London factor:
                \begin{equation}
                    \mathrm{HL} = (2J'+1)(2J''+1) \left( \begin{smallmatrix} J' & 1 & J'' \\ -\Omega' & (\Omega'-\Omega'') & \Omega'' \end{smallmatrix} \right)^2,
                \end{equation}
                where $\Omega$, once again, is the quantum number related to the projection of $J$ onto the internuclear axis, and the last term is the Wigner 3-$j$ symbol, following the convention in Ref.~\cite{brown2003rotational}.
                The initial populations of the rotational states are determined by the intricacies of the resonant-enhanced multiphoton ionization process used to create our molecular ions (see, e.g., Refs.\ \cite{zhou2019visible,ng2023thesis}). It is not a thermal distribution.
            \item Each of these discrete lines is modeled with a Gaussian profile sharing the same linewidth to be determined by the fitting algorithm. The resulting intensity profile is used for the contour fit to data.
            \item For most bands, we fix $B'' = 0.24261~\mathrm{cm}^{-1}$ and $\Omega'' = 1$, corresponding to the $X\,^3\Delta_1(v=0)$ vibronic manifold \cite{gresh2016broadband}. For bands that only appear after optical pumping has moved population between vibrational bands, we fix $B'' = 0.24161~\mathrm{cm}^{-1}$ and $\Omega'' = 1$, corresponding to the $X\,^3\Delta_1(v=1)$ vibronic manifold.
            \item Parameters allowed to vary in the fit include $T_0$, $B'$, initial population in each of the rotational states up to $J''=8$, linewidth, and overall intensity background offset. $\Omega'$ is determined by hand from the set $\lbrace 0,1,2 \rbrace$ by noting the relative intensities of the PQR branches and presence of certain lines in the band (e.g., a P(2) line would be absent for an $|\Omega|=2 \leftarrow X\,^3\Delta_1$ transition).
            \item Additionally, we denote $\Omega=0$ bands as $\Omega=0^+$ if we observe dissociation with $\Omega=0^+ \leftarrow a\,^1\Sigma^+$ as the first REMPD transition. This applies to those situations where we intentionally populate the $a\,^1\Sigma^+$ level; if the parity is not specified, the parity was not determined.
        \end{enumerate}

\begin{figure}[htb]
            \centering
            {(a) $\Omega=0$}
            \includegraphics[width=\columnwidth]{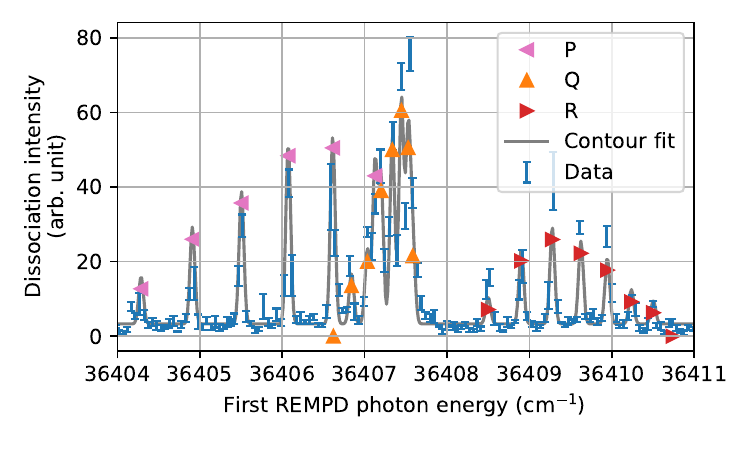} \\
            {(b) $|\Omega|=1$}
            \includegraphics[width=\columnwidth]{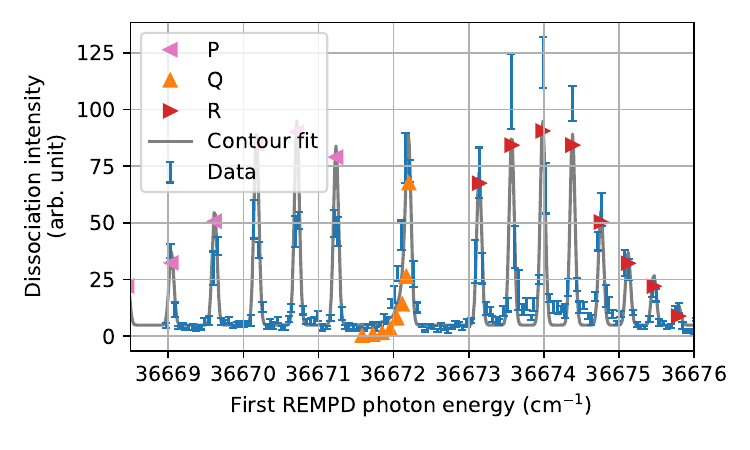} \\
            {(c) $|\Omega|=2$}
            \includegraphics[width=\columnwidth]{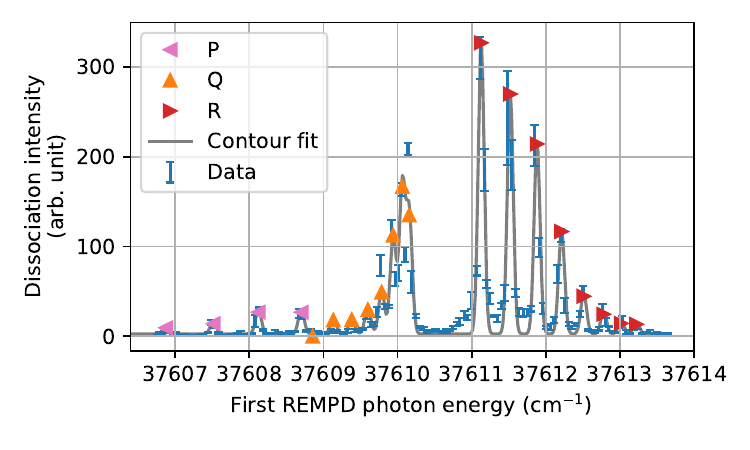}
            \caption{\textbf{Examples of contour fits performed on bands obtained from spectroscopy.} In all plots, the blue and gray traces correspond to data and contour fits, respectively. Triangle markers are used to annotate locations of the P, Q, and R lines to guide the eye. Error bars in the data are attributed to the standard error in the number of Th$^+$ ions counted across five shots. }
            \label{fig:band_example}
        \end{figure}

        The fit parameters are shown in Tables \ref{tab:bands_532} and \ref{tab:bands_355} in the appendix for bands found with 532~nm and 355~nm as the second REMPD photon, respectively. Examples of the fits are shown in Figure \ref{fig:band_example}.
        As seen in the data from the figure, we have the option of parking our laser at the desired rotational line to selectively dissociate a desired $J$ state with low levels of background from undesired $J$ states.

    \subsection{Discussion on Energy Landscape Above Dissociation Threshold}\label{sec:SS_landscape}

\begin{figure*}[t]
            \centering
            \includegraphics[width=0.95\linewidth]{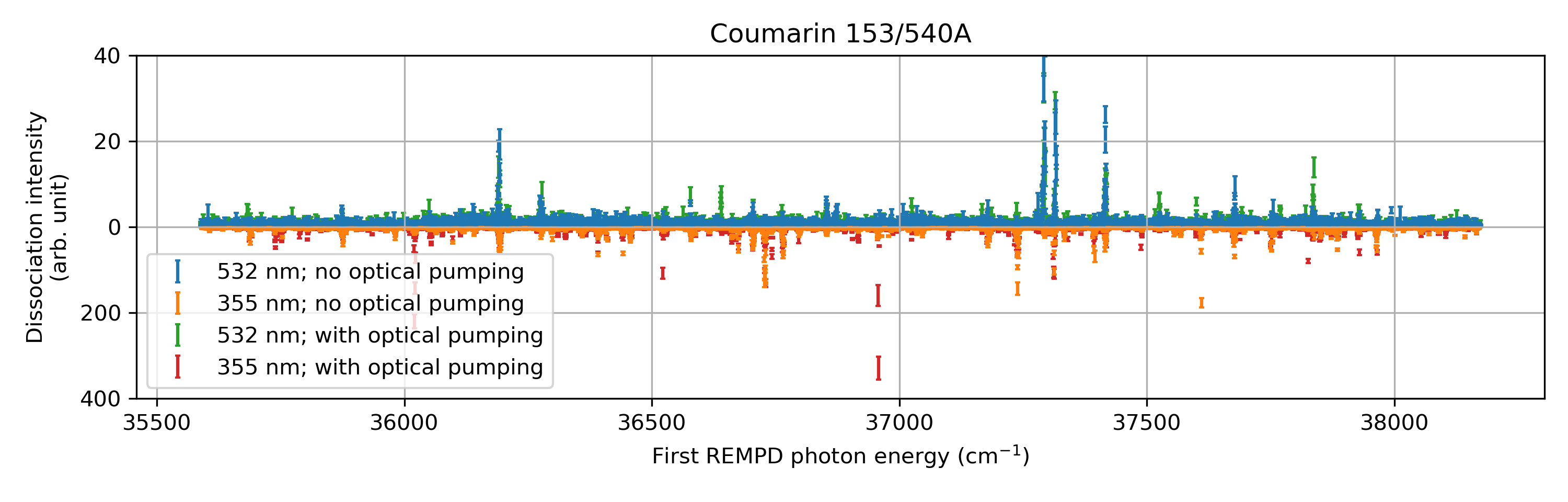} \\
            \includegraphics[width=0.95\linewidth]{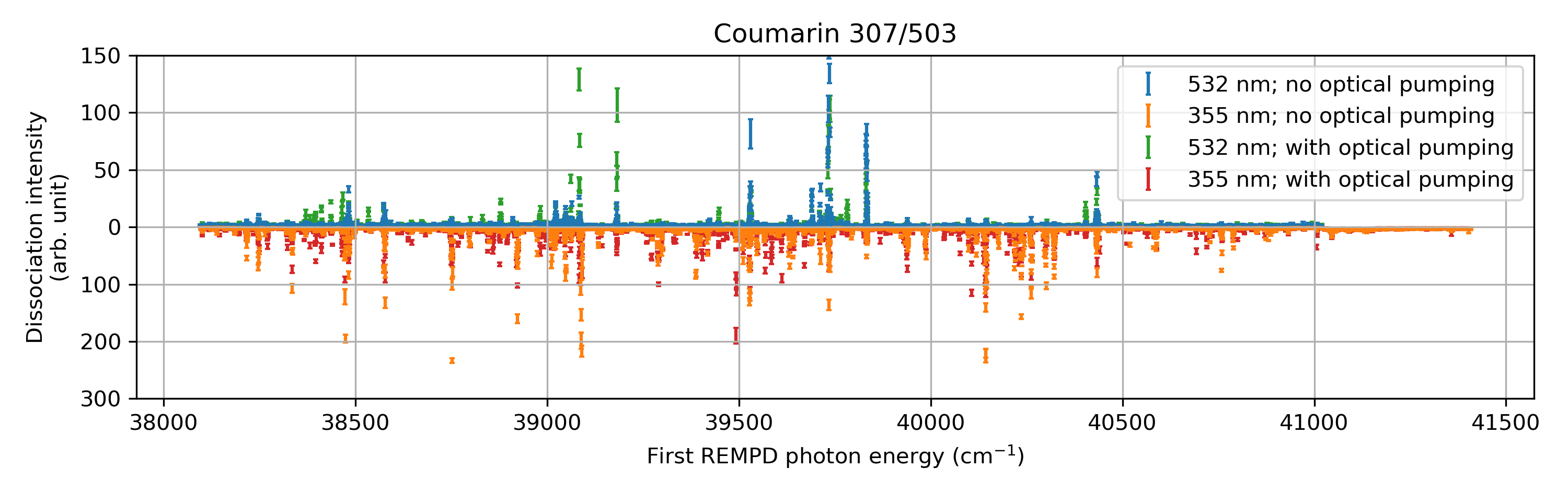}
            \caption{\textbf{Comparison of bands observed in survey spectroscopy with two different wavelengths used as the second REMPD photon.} In both plots, the bands observed with 355~nm as the second REMPD photon are presented inverted for easier comparison. For clarity, the ticks on the vertical axes are different for the 532~nm and 355~nm second REMPD photons. At this scale, each vertical line traced out by the points corresponds to a rotational band. Top: bands obtained with frequency-doubled Coumarin~153/540A as the first photon. Bottom: bands obtained with frequency-doubled Coumarin~307/503 as the first photon. The bands labeled ``no optical pumping'' and ``with optical pumping'' are attributed to excitation from the $X\,^3\Delta_1(v=0)$ and $X\,^3\Delta_1(v=1)$ vibronic manifolds, respectively.}
            \label{fig:survey_comparison}
        \end{figure*}

        Certain rotational bands are observed in the photodissociation signal only when the second photon is 532~nm, and conversely, other bands appear only when using 355~nm. This is illustrated in Figure \ref{fig:survey_comparison}.
        This observation could be attributed to the differences in the energy landscape accessed by the REMPD photons when using 532~nm versus 355~nm as the second photon. With 532~nm, we access energy levels from 1000 cm$^{-1}$ to 7000 cm$^{-1}$ above the dissociation threshold, while with 355~nm, the accessible range is higher, from 10000 cm$^{-1}$ to 16000 cm$^{-1}$. The 532~nm and 355~nm photons are illustrated in panel (d) of Figure \ref{fig:energylandscape} by the green and purple arrows, respectively.
        
        To confirm that the difference is not due to the coarse steps used in survey spectroscopy resulting in missing lines, we conduct fine scans on several strong bands and observe a distinct dependence of their intensities on the wavelength of the second REMPD photon.    Our second REMPD photon is tunable only with a single large step from 532~nm to 355~nm. It would be interesting to use a tunable laser with smaller tunable steps to characterize the structures of the energy landscape above the photodissociation threshold.

        The energy landscape of a molecule like ThF$^+$ containing a heavy atom could be quite complex, with non-adiabatic couplings of molecular potential curves adding to this complexity. This is illustrated in panel (d) of Figure \ref{fig:energylandscape}. A deeper understanding of the molecular landscape above the dissociation threshold could enable more precise quantum manipulation and more efficient quantum state readout. 

\section{Case Studies on $\Omega=2$ States}\label{sec:O2}

    We observe several strong dissociating transitions that involve intermediate states with $|\Omega| = 2$ character. These transitions exhibit dissociation efficiencies that significantly exceed those achieved in the previous REMPAD protocol used in the JILA ThF$^+$ experiment \cite{zhou2020second}. We first share results of asymmetric photofragments observed for several bands in Section \ref{sec:REMPADresults}. We then discuss state-selectivity of our REMPD protocol and the distinguishability of the two states in our two-state detection with REMPAD in Section \ref{sec:state_selectivity}. We show plots of saturation of dissociation for selected transitions in Section \ref{sec:saturation}. Finally, we discuss observed dissociation efficiency in Section \ref{sec:efficiency}.

    \subsection{Spatial Distribution of Asymmetric Photofragments}\label{sec:REMPADresults}
    
\begin{figure*}[htb]
            \centering
            \includegraphics[width=0.3\linewidth]{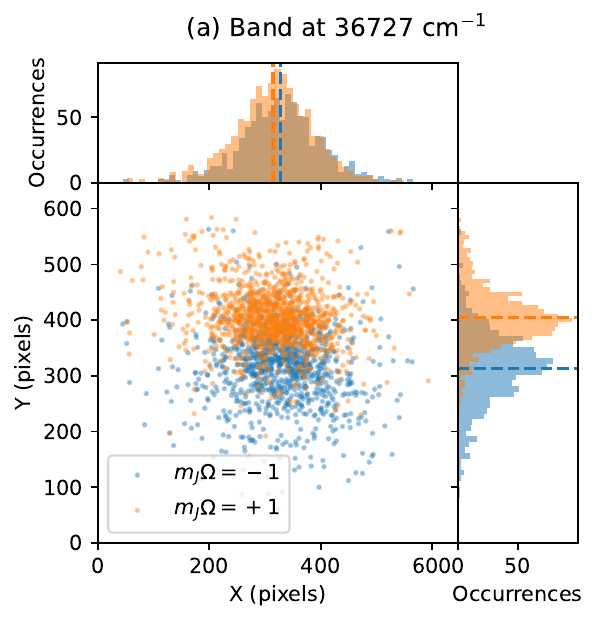}
            \includegraphics[width=0.3\linewidth]{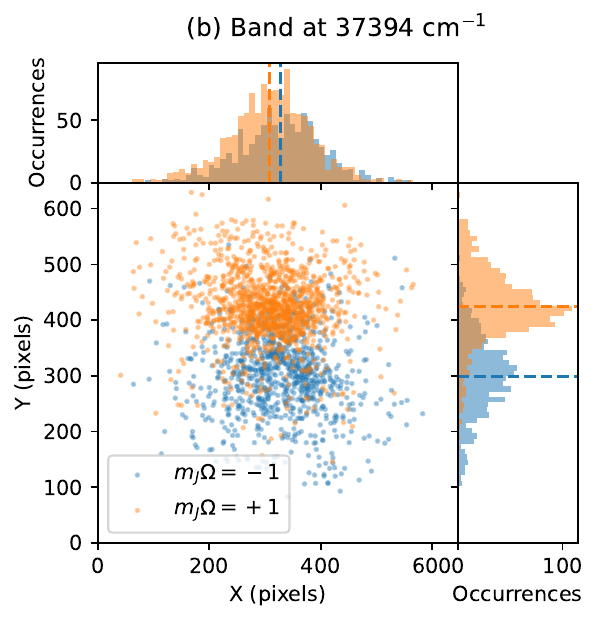}
            \includegraphics[width=0.3\linewidth]{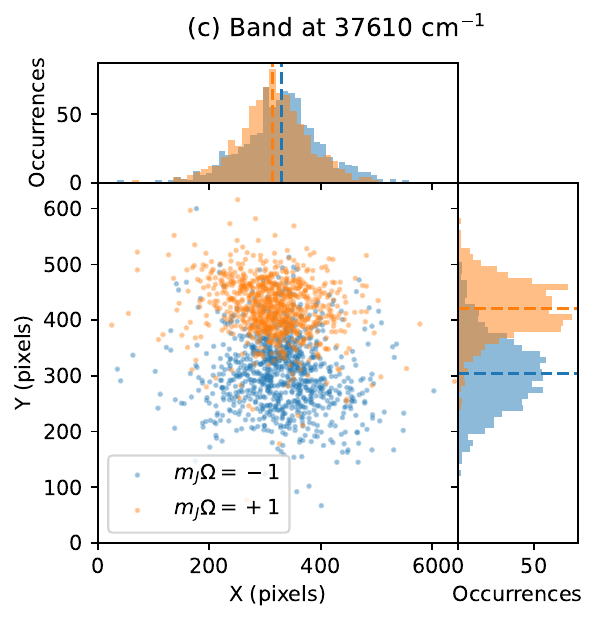}
            \caption{\textbf{Asymmetrical distribution of dissociation photofragments for three different $|\Omega|=2$ intermediate states.} Each dot in the plots represents a Th$^+$ photofragment detected by our spatially resolving ion detection system. Dissociation occurs on the R(1) line of each band, with a 355~nm photon serving as the second photon. The ThF$^+$ molecules are initially prepared in the $X\,^3\Delta_1(v=0, J=1, m_J=+1)$ state, and one of the Stark doublets (i.e., $m_J\Omega = +1$ or $m_J\Omega = -1$) is selectively depleted prior to dissociation. The dots are color-coded according to the state from which the molecules are dissociated. In each plot, data from two sets of experiment (one dissociating from $m_J\Omega=+1$ and the other from $m_J\Omega=-1$) are overlaid for comparison. Histograms displayed at the top and side of each scatter plot show the distribution of photofragments along the relevant axis, with integration along the orthogonal axis. Dashed lines indicate the first moment of the photofragment distribution, serving as a guide to the eye. The polarizing electric field is oriented in the positive vertical direction in all plots at the time of dissociation. The polarization configurations for the first and second dissociation photons in each of the bands are as follows: (a) $\sigma^+, \sigma^-$, (b) $\sigma^+, \sigma^-$, and (c) $\sigma^+, \sigma^+$.}
            \label{fig:REMPAD}
        \end{figure*}

        When dissociating on the R(1) line, we detect an asymmetric spatial distribution of the photofragments on our ion detector. This asymmetry arises from a velocity-mapped imaging of the ejection of photofragments along the molecular axis and the alignment of the molecules relative to the polarizing electric field: molecules in $m_J\Omega = +1$ states are aligned with the electric field, while $m_J\Omega = -1$ states are anti-aligned (see Figure \ref{fig:sciencestates}). Examples of these photofragment distributions are shown in Figure \ref{fig:REMPAD}. 
        At 20~V/cm, the linewidths of our REMPD lasers are broad enough to address both $m_J\Omega = \pm 1$ states at the same time. To generate the plots in Figure \ref{fig:REMPAD}, we depopulate one or the other of the $m_J\Omega = \pm 1$ states using microwaves and optical pumping \cite{ng2023thesis,zhou2020second}.
    
        We note that the spatial separation of photofragments from different $m_J\Omega$ states depends on the polarization of the second dissociation photon. This dependence may arise from the energy landscape of the molecule above the dissociation threshold \cite{ng2023thesis}. For example, maximum separation occurs when the first and second dissociation photons are $\sigma^+$ and $\sigma^-$ polarized, respectively, for the bands at 36727 and 37394~cm$^{-1}$. This suggests that these transitions connect to $|\Omega|=1$ pre-dissociation states. In contrast, the band at 37610~cm$^{-1}$ shows maximum spatial separation when both photons are $\sigma^+$ polarized, indicating that this transition connects to an $|\Omega|=3$ pre-dissociation state. For a detailed explanation, see, e.g., Chapter 3 of Ref.\ \cite{ng2023thesis}.

        We observe that the kinetic energies of photofragments for the entire range of our survey spectroscopy do not vary significantly. This suggests that while we may be connecting to pre-dissociation states with different quantum numbers, e.g., $\Omega$, at large inter-nuclear separation, these molecular states asymptotically tend to atomic states that are similar in energies.

    \subsection{State selectivity and distinguishability}\label{sec:state_selectivity}

        The key to REMPAD's state selectivity is that the angular distribution of the photofragments bears a memory of the orientation of the molecule before dissociation. Selectivity in $J$ and $v$ arises from having REMPD laser linewidths that are much narrower than the rotational spacings. As one can see from Figure \ref{fig:band_example}, the low R lines, corresponding to dissociating from low $J$ states, are very well resolved and have negligible background. 
        
        We increase the REMPD laser energies for our EDM measurements to improve dissociation efficiency, which leads to power broadening of the R lines. However, at the pulse energies achievable with our laser system, we observe that power broadening only causes slight leakage from each R line to the neighboring one. At the start of each experiment, all molecules are prepared in the $X\,^3\Delta_1(v=0,J=1)$ rovibronic manifold through optical pumping, with other rotational states (e.g., $J=2,3$) intentionally depopulated. While ion-ion collisions and blackbody radiation can excite some molecules into the $J=2$ state during the experiment, we deplete this state via optical pumping immediately before dissociation. As a result, we maintain rotational state selectivity in our REMPD scheme through dissociation on the R(1) line.

        As mentioned earlier, at a polarizing electric field strength of 20 V/cm, the linewidths of our REMPD lasers are not narrow enough to energetically resolve the Stark doublets, meaning both doublets are dissociated simultaneously. To distinguish between them, we selectively deplete the $m_J\Omega=+1$ or $-1$ states prior to dissociation (see, e.g., Figure 1 of Ref.\ \cite{ng2022spectroscopy}), leaving only one Stark doublet for analysis. This allows us to characterize how well we can differentiate whether an ion detected on one side of the ion detector corresponds to the upper or lower $m_J\Omega$ states. Despite the highly directional ejection of photofragments using our REMPAD protocol, Figure \ref{fig:REMPAD} shows that some ions from each Stark doublet leak into the other channel. This is largely due to the thermal energy of the trapped ThF$^+$ ions not being fully negligible compared to the kinetic energy of dissociation of the Th$^+$ ions. As a result, we cannot reliably assign a Stark doublet to ions that fall near the center of the ion detector. To mitigate this, we define a ``swatch'' region around the middle of the ion detector, ignoring any ions that fall within it. 
        
\begin{figure}[htb]
            \centering
            \includegraphics[width=0.49\columnwidth]{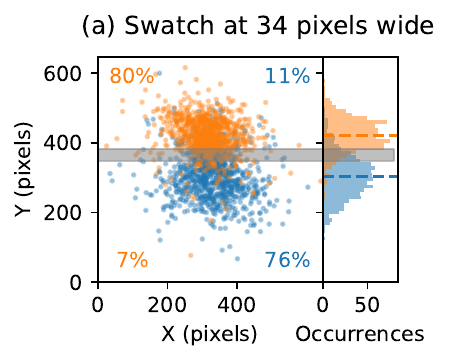}
            \includegraphics[width=0.49\columnwidth]{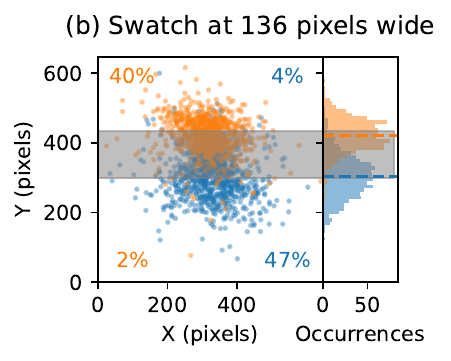}
            \caption{\textbf{Effect of swatch widths on ion counts and leakage ions.} These plots are reproduced from part (c) of Figure \ref{fig:REMPAD}. Swatches (in gray) of various widths are overlaid onto each plot. (a) Above the swatch, we count 80\% and 11\% of the ions in $m_J \Omega = +1$ and $-1$, respectively. Below the swatch, we count 7\% and 76\% of the ions in $m_J \Omega = +1$ and $-1$, respectively. (b) Above the swatch, we count 40\% and 4\% of the ions in $m_J \Omega = +1$ and $-1$, respectively. Below the swatch, we count 2\% and 47\% of the ions in $m_J \Omega = +1$ and $-1$, respectively. While contamination from the unwanted doublet has been suppressed by using a wider swatch in (b), the total ion counts also decrease. An optimal swatch width balances high ion counts with low leakage into the other channel.}
            \label{fig:swatch}
        \end{figure}

        Figure \ref{fig:swatch} illustrates an example of how the swatch can help reduce unwanted leakage into the other channel at the expense of reducing total ion counts. 
        We operate with a swatch width that balances maximal ion counts with minimal leakage. In previous work \cite{caldwell2023systematic,roussy2022thesis}, we have characterized the possible systematic effects of this leakage on the determination of the electron's EDM and have developed procedures for keeping these errors minimal. For more information on the detailed studies of the swatch using HfF$^+$, see Sections III and VII of Ref.\ \cite{caldwell2023systematic} and Section 3.8.2 of \cite{roussy2022thesis}.

    \subsection{Saturation of dissociation}\label{sec:saturation}
    
\begin{figure}[htb]
            \centering
            \includegraphics[width=\columnwidth]{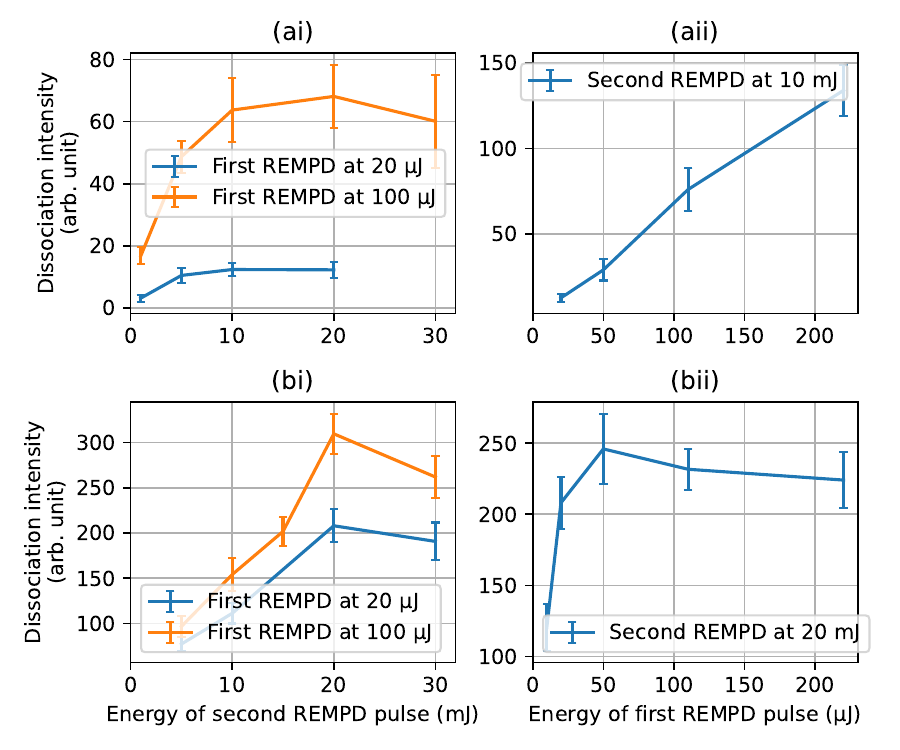}
            \caption{\textbf{Saturation of dissociated Th$^+$ ions versus REMPD laser pulse energies for selected bands using 355~nm as the second REMPD photon.} The (a) and (b) rows correspond to dissociation from the R(1) line of the bands at 37610~cm$^{-1}$ and 36727~cm$^{-1}$, respectively. The (i) column corresponds to scans performed with the first REMPD pulse energy held fixed at either 20 or 100~$\upmu$J. The (ii) column corresponds to scans performed with the second REMPD pulse energy held fixed at either 10 or 20~mJ. Saturation in dissociated ion numbers is observed within the ranges scanned in all plots except for (aii). Details of the laser beam profiles are discussed in the main text. Lines connecting the points serve to guide the eye. Each point is an average of ten shots of the experiment and error bars are associated with the standard error of dissociated ion numbers. The plotted ion yields have been corrected for off-resonant contributions of dissociation, which accounts for the decrease in effect of yields seen at high pulse energies.}
            \label{fig:saturation}
        \end{figure}

        Figure \ref{fig:saturation} contains plots showing the saturation behavior of the dissociated Th$^+$ ions as a function of REMPD laser pulse energies for selected transitions.
        Our REMPD laser beams have spatial structures due to the multi-modal nature of our pulsed lasers. Averaging over the smaller structures, we fit our REMPD laser beams to 2D Gaussian profiles with 1$\sigma$ mean radius of $\sim 4~\mathrm{mm}$ for both the first (tuned) and second (355~nm) REMPD laser pulses.
        
        The number of dissociated Th$^+$ ions saturate towards the higher pulse energies (above 20~mJ) of the second REMPD laser pulse, as can be seen from the plots in the (i) column in Figure \ref{fig:saturation}. While the results of only two selected bands are shown here, this is representative of most strong bands found in our survey scan. 
        
        Similarly, the dissociated number saturates beyond 50~$\upmu$J of the first REMPD laser pulse for the band at 36727~cm$^{-1}$, as can be seen in part (bii) of Figure \ref{fig:saturation}. However, this is not representative of most strong bands found in our survey scan. Instead, most of the bands exhibit saturation behavior, or the lack thereof, similar to that shown in part (aii) in Figure \ref{fig:saturation}, where the dissociated ion numbers continue to increase linearly with increasing first REMPD laser pulse energy.

    \subsection{Dissociation and detection efficiency}\label{sec:efficiency}

        For dissociation efficiency studies, we use the highest first REMPD laser pulse energies achievable with our setup, limited by the output of our dye laser. We keep the second REMPD laser pulse energy at 30~mJ.

        We determine the dissociation efficiencies of various transitions as follows. First, we scan a band with well-separated lines (e.g., the band at 37610~cm$^{-1}$) to deduce the relative population of molecules in each rotational state. We observe that the distribution of molecules across rotational states remains stable from day to day and we characterize the level of fluctuation. Second, we dissociate only the $J = 1$ state via the R(1) line of the band of interest, simultaneously counting the dissociated Th$^+$ ions and the undissociated ThF$^+$ molecules with our ion detector. Finally, we extract the dissociation efficiency by taking the ratio of the Th$^+$ counts to the fractional population of ThF$^+$ in the $J = 1$ state. We ensure that our Th$^+$ and ThF$^+$ counts are not affected by saturation effects in the ion detector. To the best of our ability, we ensure that the entire ion cloud is illuminated by both dissociation laser pulses. 
              
        We observe dissociation efficiency as high as 57(14)\% on the R(1) line in the transition with the band at 36727~cm$^{-1}$, more than ten times as efficient as the transition used in our previous protocol \cite{ng2022spectroscopy, zhou2019visible}. The uncertainty is associated with the combined standard deviations from fluctuations in the rotational population distribution and detected ion counts. We notice that depleting the $J=2$ state before dissociation changes the number of detected dissociated Th$^+$ ions by less than about 2\%, suggesting that leakage from neighboring R lines due to power broadening is minimal. The state selectivity for the band at 36727~cm$^{-1}$ shown in Figure~\ref{fig:REMPAD} is expected to improve as the ion temperature is reduced below the current level ($\sim$30 K), an improvement that is the focus of ongoing efforts in the JILA experiment.

        Most of the observed transitions exhibit dissociation efficiencies on the order of 30\%, also surpassing that used in our previous protocol. Since we are limited by the pulse energies from our lasers, we do not make attempts to make more systematic studies on increasing dissociation efficiencies beyond what are currently observed.

        The overall detection efficiency is more involved. This involves the following: (i) dissociation efficiency, (ii) ion kickout transfer efficiency, and (iii) detector efficiency. The first has already been discussed. The second depends on a multitude of factors. After dissociation, all ions are kicked towards the ion detector. To optimize the kickout transfer efficiency, we adjust the ion trajectory by steering the ions during kickout and tuning the ion optics to modify the focusing strength. We aim to keep the ions well-focused, ensuring that most reach the detector, but slightly defocused to avoid ion pile-up and allow the ions to spread across the detector surface for counting via imaging. We estimate the fraction of ions that miss the detector to be less than 10\%. A larger detector and better controlled kickout conditions will further suppress ion loss. The third involves an interplay of the fractional active area and gain of the microchannel plates. The open area ratio is about 70\% in our microchannel plate assembly. The impact energy of the ThF$^+$ ions is limited to approximately 3~keV, at which the quantum efficiency is estimated to be 30\% \cite{krems2005channel}. Overall, we estimate the detection efficiency to be $57(14)\% \times 90\% \times 70\% \times 30\% \approx 11\%$.

\section{Prospects of Using $\Omega=0^\pm$ States for State Detection}\label{sec:O0}

    REMPAD has allowed us to simultaneously measure the population in two distinct molecular levels. The associated suppression of various sources of common-mode technical noise was a key advance responsible for much of the improvement in precision between the first \cite{cairncross2017precision} and second \cite{Roussy2023an} generations of the JILA electron's EDM measurement. 
    
    A significant advance would be the ability to separately measure the populations in all four of the states shown in Figure \ref{fig:sciencestates}. Conceptually, one could convert from spatially resolved detection of dissociation products to temporally resolved. One could, for instance, apply four sets of REMPD laser pulses, separated by a few milliseconds in time. The four sets of pulses would represent different combinations of energies and helicities. Resonance conditions and angular-momentum selection rules would cause the four sets of laser pulses to each dissociate a distinct state of the four shown in Figure \ref{fig:sciencestates}. 
    \begin{figure}[htb]
        \centering
        \includegraphics[width=\columnwidth]{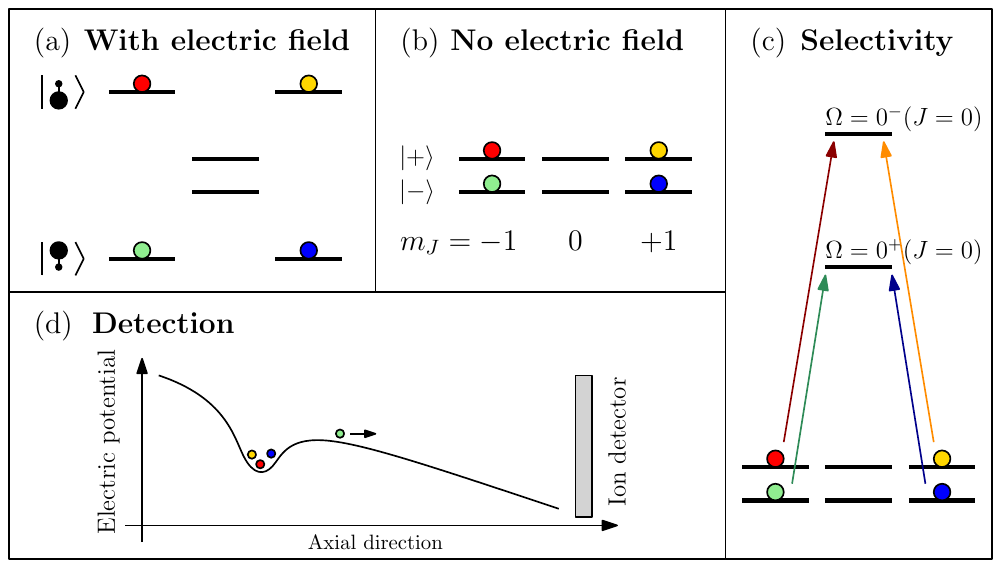}
        \caption{\textbf{Proposed four-state detection scheme.} (a) Energy structure of the $J=1$ manifold of the $X\,^3\Delta_1(v=0)$ ``science state''; a redraw of Figure \ref{fig:sciencestates}. Following a Ramsey sequence, we want to measure the four populations indicated by disks colored red, yellow, green and blue. Not shown are the hyperfine structure and the effects of the magnetic field. (b) The $J=1$ manifold after the bias electric field has been adiabatically turned off. In a simplified understanding, the color-coded populations shown in (a) track in to the corresponding colors in (b). (c) Schematic level diagram showing the $X\,^3\Delta_1(v=0,J=1)$ levels again and two $\Omega=0$ states of opposite parity. The $\Omega=0$ states are spaced by $\sim100~\mathrm{cm}^{-1}$ and are located $\sim30000~\mathrm{cm}^{-1}$ above the $X\,^3\Delta_1$ ground state. Four sets of REMPD laser pulses with varying energies and helicities are indicated. Each set of laser pulses is separated from the others by temporal delays of tens of milliseconds. Energy resonance conditions allow for coupling only to the $J=0$ excited states (a P(1) transition). Parity and angular momentum selection rules are such that the populations represented by each of the colored disks can be dissociated by only one of the laser pulses. (d) Schematic electrostatic potential showing the confining potential trap plus an approximately linear gradient that will accelerate the photodissociated products (which escape from the trap) onto the ion detector.}
        \label{fig:AdiabaticMapping}
    \end{figure}
    The corresponding dissociation product would arrive at our detector spaced by the temporal delays between the sets of pulses. Unfortunately, as a practical matter, the upper and lower pairs of states in Figure \ref{fig:sciencestates} are separated by only $\sim100~\mathrm{MHz}$ in ThF$^+$, not enough to resolve with our lasers with high contrast.

    Our extensive survey spectroscopy of intermediate states for REMPD has pointed us to a way around the energy-resolution limitation. The procedure we propose below is still under development, but could offer several advantages. In our survey data, we found a number of $\Omega=0$ states with good photodissociation efficiency and well-defined parity. In several cases the odd and even parity $\Omega=0$ states are spaced by $\sim100~\mathrm{cm}^{-1}$, readily resolved by our pulsed lasers. In its simplest version, a four-state detection protocol could go as follows.
    \begin{enumerate}
        \item As shown in panel (a) of Figure \ref{fig:AdiabaticMapping}, after the Ramsey sequence, we have varying populations of molecules in four levels of the $X\,^3\Delta_1(v=0,J=1)$ manifold, states characterized by good alignment of the molecule along a bias electric field.
        \item Adiabatically ramp down the bias electric field so that the levels are now states of good parity, separated by the $\Omega$-doubling splitting of $\sim10~\mathrm{MHz}$. For now, let us assume that the populations distinguished by color in panel (a) of Figure \ref{fig:AdiabaticMapping} adiabatically track to the same colors in panel (b).
        \item Apply four sets of REMPD laser pulses, each spaced by a few tens of milliseconds of time delay, with the relevant helicities and energies such as to be resonant with a $\Omega=0^+(J=0)$ or $\Omega=0^-(J=0)$ level.
        \item As shown in panel (d) of Figure \ref{fig:AdiabaticMapping}, the ions are confined in a shallow ion trap superimposed on a linear gradient, which can accelerate ions towards a detector. The recoil energy of photodissociation would be greater than the trap depth, so that as each level is dissociated, the Th$^+$ ions fall on the ion detector and are separately counted. The concept here is basically that of trap-loss spectroscopy \cite{wang2007ultra} and similar to leak-out spectroscopy \cite{schmid2022leak}.
    \end{enumerate}
    
    The vision presented above is oversimplified. In particular, in the presence of (i) rotation of the bias electric field, (ii) a magnetic field, and (iii) hyperfine structure, the effect of adiabatically ramping down the electric field is less ideal than is suggested in panels (a) and (b) of Figure \ref{fig:AdiabaticMapping}. In reality, ensuring that the populations wind up in the desired final states will require engineered ramps and pulse sequences. We are currently exploring these issues. 
    
    While the vision presented above is inherently interesting and may have applications in molecular physics or quantum information, the potential gain in data collection over the current protocol is, at best, a factor of two. Moreover, several technical challenges remain unresolved, and it is unclear whether even this modest improvement can be achieved in practice. Therefore, we consider it unlikely that this approach will be adopted in a fully developed EDM experiment.
        
\section{Conclusion}\label{sec:conclusion}

    We conducted survey spectroscopy from 35500~cm$^{-1}$ to 41500~cm$^{-1}$ above the ground state to identify intermediate states with high dissociation efficiency. Several intermediate states were found to exhibit dissociation efficiencies surpassing those used in our previous protocol. Molecular parameters for these states were extracted through contour fitting. Among the identified bands, several were found to have favorable quantum numbers, specifically $|\Omega| = 2$, suitable for two-state detection using REMPAD. We also propose a new protocol based on adiabatic mapping, which explores the potential of using $\Omega = 0^\pm$ states for multi-state detection. While this work focuses on ThF$^+$, we believe the principles of REMPAD and adiabatic mapping could be extended to other molecular ions with similar structures.

\section*{Acknowledgment}
This work is supported by the Natural Sciences and Engineering Council of Canada (NSERC), Moore Foundation, Sloan Foundation, NSF PFC (PHY 2317149), NIST, AFOSR, Marsico Research Chair, and the National Science Foundation, under Grant No.\ PHY-2309253 (L.\ C.). TRIUMF receives federal funding via a contribution agreement with the National Research Council of Canada. K.~B.~Ng acknowledges support from the Banting Postdoctoral Fellowship BPF-198564. We thank X.~Fan for comments on the manuscript.

\bibliography{biblio}

\appendix
\section{Identified Bands}

    We perform contour fits for 12 and 45 bands identified with 532~nm and 355~nm as the second REMPD photon, respectively. The fitting procedure is detailed in Section \ref{sec:SS_fitting}. Fit parameters are presented in Tables \ref{tab:bands_532} and \ref{tab:bands_355} for the bands associated with 532~nm and 355~nm second REMPD photon, respectively. 

    \begin{table*}[htb]
        \centering
        \rowcolors{2}{gray!20}{white}
        \begin{tabular}{l l l l l}
        \toprule
            {$T_0$ (cm$^{-1}$)} & {$B'$ (cm$^{-1}$)} & {$\Omega'$} & {$v''$} & {Additional notes} \\
        \midrule
            36191.0 & 0.215(?) & 2 & 0 & Could correspond to the 36190 band in Table \ref{tab:bands_355}. \\
            37292.1 & 0.22237(5) & 2 & 0 & - \\
            37315.9 & 0.20467(4) & 0 & 0 & Distinctly different band from 37313 band in Table \ref{tab:bands_355}. \\
            37416.4 & 0.217(?) & 0 & 0 & Could correspond to the 37416 band in Table \ref{tab:bands_355}. \\
            37835.9 & 0.2217(3) & 1 & 1 & - \\
            39091.4 & 0.239(?) & 0 & 1 & Could correspond to the 39091 band in Table \ref{tab:bands_355}. Possible $(v',1'')$ partner to 39744. \\
            39189.5 & 0.241(?) & 0 & 1 & - \\
            39539.1 & 0.2211(8) & 0 & 0 & - \\
            39744.5 & 0.2325(3) & 0 & 0 & Could correspond to the 39744 band in Table \ref{tab:bands_355}. Possible $(v',0'')$ partner to 39091. \\
            39842.7 & 0.22406(6) & 0 & 0 & Could correspond to the 39842 band in Table \ref{tab:bands_355}. \\
            40415.4 & 0.237(?) & 1 & 1 & - \\
            40444.7 & 0.23737(9) & 0 & 0 & - \\
        \bottomrule
        \end{tabular}
        \caption{\textbf{Bands identified with 532~nm as the second photon.} All $v''=0$ and $v''=1$ bands are fit to $B'' = 0.24261~\mathrm{cm}^{-1}$ and $B'' = 0.24161~\mathrm{cm}^{-1}$, respectively. Uncertainties are attributed to 90\% confidence level obtained from the bootstrap procedure \cite{efron1994introduction} with 100 iterations. We observe long term drifts in the wavemeter frequency calibration of about 0.1~cm$^{-1}$, so all $T_0$ values are reported with an uncertainty of this value, much larger than the uncertainties obtained from the bootstrap procedure. The raw data for bands marked with uncertainties labeled as ``(?)'' fit poorly with the fit function due to several factors, including low signal-to-noise ratio, presence of overlapping bands, and the presence of unidentified structures. For these bands, $B'$ is reported to 3 decimal places. We use combination differences on bands with the same $\Omega'$ to infer vibrational progressions on the $X\,^3\Delta_1 (v''=0,1)$ state. We assume that the transitions address the same $v'$ in the excited state. We denote these correlations in the table with $(v',v'')$, where $v''\in\lbrace 0,1\rbrace$.}
        \label{tab:bands_532}
    \end{table*}
    
    \begin{table*}[htb]
        \centering
        \rowcolors{2}{gray!20}{white}
        \begin{tabular}{l l l l l}
        \toprule
            {$T_0$ (cm$^{-1}$)} & {$B'$ (cm$^{-1}$)} & {$\Omega'$} & {$v''$} & {Additional notes} \\
        \midrule
            36019.1 & 0.2320(17) & 1 & 1 & Possible $(v',1'')$ partner to 36672. \\
            36190.9 & 0.223(?) & 2 & 0 & Could correspond to the 36190 band in Table \ref{tab:bands_532}. \\
            36390.3 & 0.2281(2) & 0 & 0 & - \\
            36407.6 & 0.22885(15) & 0 & 0 & - \\
            36440.4 & 0.2241(3) & 0 & 0 & - \\
            36520.7 & 0.2246(8) & 1 & 1 & - \\
            36672.2 & 0.23372(13) & 1 & 0 & Possible $(v',0'')$ partner to 36019. \\
            36727.1 & 0.2276(6) & 2 & 0 & - \\
            36740.9 & 0.2284(13) & 2 & 1 & Possible $(v',1'')$ partner to 37394. \\
            36764.5 & 0.2158(4) & 0 & 0 & - \\
            36957.1 & 0.224(2) & 2 & 1 & Possible $(v',1'')$ partner to 37610. \\
            37239.0 & 0.2251(10) & 2 & 0 & - \\
            37313.4 & 0.2161(4) & 0$^+$ & 0 & Distinctly different band from 37315 band in Table \ref{tab:bands_532}. \\
            37394.0 & 0.22868(7) & 2 & 0 & Possible $(v',0'')$ partner to 36740. \\
            37416.4 & 0.220(?) & 0 & 0 & Could correspond to the 37416 band in Table \ref{tab:bands_532}. \\
            37610.3 & 0.22711(9) & 2 & 0 & Possible $(v',0'')$ partner to 36957. \\
            37753.3 & 0.2137(2) & 0 & 0 & - \\
            37826.7 & 0.227(?) & 0 & 1 & Possible $(v',1'')$ partner to 38479. \\
            37886.4 & 0.2246(6) & 0$^+$ & 0 & - \\
            37964.4 & 0.2248(2) & 2 & 0 & - \\
            38253.2 & 0.226(3) & 0 & 0 & - \\
            38341.7 & 0.22570(11) & 0 & 0 & - \\
            38402.2 & 0.218(?) & 2 & 1 & Possible $(v',1'')$ partner to 39055. \\
            38479.9 & 0.2212(7) & 0 & 0 & Possible $(v',0'')$ partner to 37826. \\
            38489.0 & 0.2216(10) & 2 & 0 & - \\
            38581.0 & 0.218(2) & 0 & 0 & - \\
            38584.4 & 0.2187(14) & 0 & 0 & - \\
            38759.1 & 0.223(3) & 0 & 0 & - \\
            38804.8 & 0.2127(2) & 0 & 0 & - \\
            38929.9 & 0.2243(14) & 0 & 0 & - \\
            39055.3 & 0.2169(7) & 2 & 0 & Possible $(v',0'')$ partner to 38402. \\
            39091.4 & 0.239(?) & 0 & 1 & Could correspond to the 39091 band in Table \ref{tab:bands_532}. Possible $(v',1'')$ partner to 39744. \\
            39097.1 & 0.23143(14) & 1 & 0 & - \\
            39395.8 & 0.23061(15) & 2 & 0 & - \\
            39501.4 & 0.217(3) & 0 & 1 & Possible $(v',1'')$ partner to 40154. \\
            39620.2 & 0.220(?) & 0 & 1 & - \\
            39721.2 & 0.2296(2) & 1 & 0 & - \\
            39744.6 & 0.225(?) & 0 & 0 & Could correspond to the 39744 band in Table \ref{tab:bands_532}. Possible $(v',1'')$ partner to 39091. \\
            39842.6 & 0.2239(5) & 0 & 0 & Could correspond to the 39842 band in Table \ref{tab:bands_532}. \\
            40117.3 & 0.230(3) & 0 & 1 & - \\
            40154.4 & 0.2243(13) & 0 & 0 & Possible $(v',0'')$ partner to 39501. \\
            40246.8 & 0.2256(6) & 0 & 0 & - \\
            40311.8 & 0.23175(17) & 1 & 0 & - \\
            40332.9 & 0.2304(2) & 0 & 0 & - \\
            40529.9 & 0.2259(4) & 2 & 0 & - \\
        \bottomrule
        \end{tabular}
        \caption{\textbf{Bands identified with 355~nm as the second photon.} All $v''=0$ and $v''=1$ bands are fit to $B'' = 0.24261~\mathrm{cm}^{-1}$ and $B'' = 0.24161~\mathrm{cm}^{-1}$, respectively. Uncertainties are attributed to 90\% confidence level obtained from the bootstrap procedure \cite{efron1994introduction} with 100 iterations. We observe long term drifts in the wavemeter frequency calibration of about 0.1~cm$^{-1}$, so all $T_0$ values are reported with an uncertainty of this value, much larger than the uncertainties obtained from the bootstrap procedure. The raw data for bands marked with uncertainties labeled as ``(?)'' fit poorly with the fit function due to several factors, including low signal-to-noise ratio, presence of overlapping bands, and the presence of unidentified structures. For these bands, $B'$ is reported to 3 decimal places. We use combination differences on bands with the same $\Omega'$ to infer vibrational progressions on the $X\,^3\Delta_1 (v''=0,1)$ state. We assume that the transitions address the same $v'$ in the excited state. We denote these correlations in the table with $(v',v'')$, where $v''\in\lbrace 0,1\rbrace$.}
        \label{tab:bands_355}
    \end{table*}

\end{document}